\documentstyle[preprint,aps]{revtex}
\newcommand{\be}{\begin{eqnarray}}
\newcommand{\ee}{\end{eqnarray}}

\begin{document}
\draft
\title{Nuclear Effects in Deep Inelastic
Scattering of Polarized Electrons off Polarized $^3$He
and the Neutron Spin Structure Functions}
\author{C.Ciofi degli Atti and S. Scopetta\cite{scop}}
\address{
Dipartimento di Fisica,
Universit\`a di Perugia and Istituto Nazionale di Fisica Nucleare,
Sezione di Perugia\\
Via A.Pascoli, I-06100 Perugia, Italy \\
}
\author{E. Pace}
\address{Dipartimento di Fisica, Universit\`a di Roma ``Tor Vergata",
and Istituto Nazionale di Fisica Nucleare, Sezione Tor Vergata, Via
E. Carnevale, I-00173 Roma, Italy}
\author{G. Salm\`e}
\address{Istituto Nazionale di Fisica Nucleare, Sezione Sanit\`a,
V.le Regina Elena 299, I-00161 Roma, Italy}
\maketitle

\begin{abstract}
It is shown that the nuclear effects playing a relevant role in
Deep Inelastic Scattering of polarized electrons by polarized $^3$He
are mainly those arising from the effective proton and neutron
polarizations generated by the $S'$ and $D$ waves in $^3$He. A simple
and reliable equation relating the neutron, $g_1^n$, and $^3$He,
$g_1^3$, spin structure functions is proposed. It is shown that the
measurement of the first moment of the $^3$He structure function can provide
a significant check of the Bjorken Sum Rule.
\end{abstract}

\narrowtext
The spin structure functions (SSF) of the nucleon $g_1^N$ and
$g_2^N$ provide information on the spin distribution among the
nucleon partons and can allow
important tests of various models of hadron's structure \cite{gen}. Our
experimental knowledge is limited at present to the proton SSF $g_1^p$
\cite{slac1,emc} and it is for this reason that new experiments
\cite{smc,slac2,herm} are under way aimed at improving the knowledge
of $g_1^p$, as well as
at measuring, for the first time, the proton SSF $g_2^p$ and the
neutron SSF $g_1^n$ and $g_2^n$.
The latter quantities are expected to be obtained from the
spin
asymmetry
measured in Deep Inelastic Scattering (DIS) of longitudinally
polarized electrons off polarized nuclear targets, $viz$ $\vec{^2{\rm H}}$
and $\vec{^3{\rm He}}$. As is well known, the use
of $\vec{^3{\rm He}}$ targets, which will be considered in this paper,
is motivated by the observation that,
in the simplest picture of $^3{\rm He}$
(all nucleons in $S$ wave),
protons have
opposite spins so that their
contribution
to the asymmetry largely cancels out. However, such a cancellation
does not occur
if other components
of the three body wave function are considered;
moreover, the fact that electrons scatter
off nucleons having a certain momentum and energy distribution
may, in principle, limit the possibility to obtain information
on nucleon SSF from scattering experiments on nuclear targets.
The aim of this Rapid
Communication is to quantitatively
illustrate whether and to which extent the extraction of $g_1^n$
from the asymmetry of the process
$\vec {^3{\rm He}}(\vec e, e')$X could be
hindered by nuclear effects arising
from small wave function components of $^3$He, as well as from
Fermi motion and binding correction
effects on DIS. To this end, we use
the spin dependent spectral
function of $^3$He \cite{cps1}, which allows one to take into account
at the same time Fermi motion and binding corrections, unlike previous
calculations \cite{wol} where only Fermi
motion effects were considered.
It should be pointed out that our paper is also based on a
recent, improved description of inclusive scattering
of polarized electrons by polarized nuclei
\cite{sa,cps2}, which leads in
the quasi elastic kinematics to appreciable
differences with respect to previous calculations
\cite{cps1,Bla}; therefore we will
also check whether these
differences persist in the DIS region.

In the Bjorken limit the longitudinal asymmetry for in\-clu\-si\-ve scattering
of longitudinally polarized electrons off a
polarized $J={1\over2}$ target with atomic
weight $A$,
reads as follows:
\be
A_{||}={\sigma_{\uparrow \uparrow} - \sigma_{\uparrow \downarrow}
\over \sigma_{\uparrow \uparrow} + \sigma_{\uparrow \downarrow}}=
2x{g_1^A(x) \over F_2^A(x)} \equiv A_{\vec A} \label{as}
\ee
where $\sigma_{\uparrow \uparrow (\uparrow \downarrow)}$ is the
differential cross section corresponding to the target spin parallel
(antiparallel) to the electron spin, $x=Q^2/ 2M\nu$ is the Bjorken
variable, $g_1^A$ and $F_2^A$ are
the nuclear spin-dependent and spin--independent
structure functions of the target A.
In what follows, three models for the asymmetry,
in order of increasing complexity, will be considered, $viz$:
\\
1) {\it No nuclear effects}. This model is such that the following
equations hold:
\be
g_1^3(x) & = & g_1^n(x) \label{gmod1} \\
A_{\vec {^3{\rm He}}} & = & f_n A_{\vec n} \label{mod1}
\ee
where $A_{\vec n}(x)=2x g_1^n(x) /F_2^n(x)$ is the neutron asymmetry and
$f_n=F_2^n(x)/(2F_2^p(x) + F_2^n(x))$ the neutron
dilution factor. Such a picture
is equivalent to consider polarized electron scattering off
$\vec {^3{\rm He}}$ described as a
pure symmetric $S$ wave
disregarding, moreover, Fermi motion and binding effects.
\\
2) {\it Proton contribution within realistic wave function of $^3$He.}
Besides the $S$ wave, the three body wave function contains a
percentage of $S'$ and $D$ waves,
$P_{S'}$ and $P_D$,
which are responsible for a proton
contribution to the polarization of $\vec{^3{\rm He}}$. The amount of such a
contribution can be calculated by
considering the quantities $P_{p(n)}^{(\pm)}$,
representing the probability
to have a proton (neutron) with spin parallel (+) or antiparallel (--)
to $^3$He spin. In a pure $S$ wave state $P_n^{(+)}=1$, $P_n^{(-)}=0$ and
$P_p^{(+)}=P_p^{(-)}={1\over2}$, whereas for a three--body wave function
containing $S$, $S'$ and $D$ waves, one has
\cite{fri,kap}
$P_n^{(+)}=1-\Delta$, $P_n^{(-)}=\Delta$, $P_p^{(\pm)}={1 \over 2}
\mp \Delta'$, where $\Delta={1 \over 3} [P_{S'}+2P_D]$ and
$\Delta'={1 \over 6}[P_D-P_{S'}]$.
{}From
world calculations on the three body system one obtains,
in correspondence of the experimental value of the
binding energy of $^3$He,
$\Delta=0.07 \pm 0.01$ and $\Delta'=
0.014\pm 0.002$ \cite{fri}.
If the $S'$ and $D$ waves are
considered and
Fermi motion and binding effects
are disregarded, one can write:
\be
g_1^3(x) & = & 2p_p g_1^p(x) + p_n g_1^n(x)\label{gmod2} \\
A_{\vec {^3{\rm He}}}&  = & 2 f_p p_p A_{\vec p} + f_n p_n A_{\vec n}
\label{mod2}
\ee
where $f_{p(n)}=
F_2^{p(n)}/(2F_2^p+F_2^n)$
is the proton (neutron) dilution factor,
$A_{{\vec p}(\vec n)}=2x g_1^{p(n)}/F_2^{p(n)}$
is the proton (neutron) asymmetry
and the effective nucleon polarizations are:
\be
p_p & = & P_p^{(+)}-P_p^{(-)}=-0.028{\pm} 0.004
\label{polp} \\
p_n & = & P_n^{(+)}-P_n^{(-)}=0.86 {\pm} 0.02 \label{pol}
\ee
3) {\it Proton contribution within the convolution approach}.
In order to take into account Fermi motion and binding effects, we
have extended to polarized DIS the usual convolution approach
adopted to treat the unpolarized DIS \cite{cps3}.
Let
us first consider the general case of inclusive scattering by spin
${1 \over 2}$ targets in impulse
approximation. We obtain for
the nuclear spin structure function $g_1^A$ the following
expression:
\widetext
\be
g_1^A(x,Q^2) & = & \sum_N \int \! dz
\int \! dE \int \! d{\bf p}
\Bigg \{ { 1 \over z }g_1^N
\left( {x \over z},Q^2 \right)  \Bigg[ P_{||}^N({\bf p},E)
+ \left( {p_{||} \over E_p+M} -{\nu \over |{\bf q}|} \right)
{|{\bf p}|\over M}{\cal P}^N ({\bf p},E)\Bigg] \Bigg. \nonumber \\
&  & - C {Q^2 \over |{\bf q}|^2} { 1 \over z }g_1^N
\left( {x \over z},Q^2 \right)
{1 \over M}
\Bigg[ {|{\bf p}|^2 \over 2(E_p+M)}\Phi(\alpha){\cal P}^N({\bf p,}E)+
{M \over 2} {\Phi (\alpha) \over \sin \alpha} P_{\bot}^N({\bf p},E)
\Bigg] \Bigg. \nonumber \\
&  & + C  {Q^2 \over |{\bf q}|^2}{1 \over z^2} g_2^N
\left( {x \over z},Q^2 \right)
{1 \over M}
\Bigg[ {|{\bf p}|^2 \over 2(E_p+M)}\Phi(\alpha){\cal P}^N({\bf p},E)-
{E_p\over 2} {\Phi(\alpha) \over \sin \alpha} P_{\bot}^N({\bf p},E)\Bigg.
\nonumber \\
&  &
-{ |{\bf q}| \over \nu} p_{||} \left( P_{||}^N({\bf p},E)-
{P_{\bot}^N({\bf p},E) \over \tan \alpha} \right) \Bigg ]
\Bigg \}
\delta \left( z-{p \cdot q \over M \nu} \right) \label{an}
\ee
\narrowtext
where $p\equiv(p^0,{\bf p})$ is the four--momentum of the bound nucleon,
with $p^0=M_A-\left[(E-M+M_A)^2+|{\bf p}|^2 \right]^{1 \over 2}$;
$E$ is the nucleon removal energy;
$E_p=\left[
M^2 + |{\bf p}|^2 \right]^{1 \over 2}$;
$\Phi(\alpha)=(3 \cos^2 \alpha -1)/\cos \alpha$,
with $\cos \alpha = {\bf p} \cdot {\bf q} /|{\bf p}||{\bf q}|$;
$p_{||}=|{\bf p}| \cos\alpha$;
$C$ is a constant
to be discussed later on; $P_{||}^N({\bf p},E)$, $P_{\bot}^N({\bf p},E)$,
${\cal P}^N({\bf p},E)$ are defined as follows
\cite{cps1}:
\be
P_{||}^N ({\bf p},E)& = & P^N_{{1 \over 2}{1\over 2}M}({\bf p},E)-
P^N_{-{1 \over 2}-{1\over 2}M}({\bf p},E) \label{par} \\
P_{\bot}^N({\bf p},E) & = & 2 P^N_{{1 \over 2}-{1\over 2}M}({\bf p},E)
e^{i \phi}, \label{ort} \\
{\cal P}^N ({\bf p},E)& = & \sin \alpha P_{\bot}^N({\bf p},E)+
\cos \alpha P_{||}^N({\bf p},E) \label{Pi} \label{cors}
\ee
where $\phi$ is the polar angle, and
\be
P_{\sigma \sigma'M}^N ({\bf p},E) & = & \sum_f[\langle
\psi^f_{A-1};N,{\bf p},\sigma'|\psi_{J,M} \rangle]^* \nonumber \\
& & [\langle
\psi^f_{A-1};N,{\bf p},\sigma|\psi_{J,M} \rangle] \nonumber \\
& & \delta(E-E_{A-1}^f+E_A)~~. \label{spet}
\ee
is the spin dependent spectral function.
Of particular relevance are the ``up" and ``down" spectral functions
$P^N_{{1 \over 2}{1 \over 2}{1 \over 2}}$ and $P^N_{-{1 \over 2}-{1 \over 2}
{1 \over 2}}$, respectively,
for they determine the effective nucleon polarization,
$viz$:
\be
P_N^{(+)} & = & \int P^N_{{1 \over 2}{1 \over 2}{1 \over 2}} ({\bf p},E)
d {\bf p} dE
\label{up} \\
P_N^{(-)} & = & \int P^N_{-{1 \over 2}-{1 \over 2}{1 \over 2}} ({\bf p},E)
d {\bf p} dE~~. \label{down}
\ee
Using in Eq.\ (\ref{an}) the proper nucleon SSF $g_{1(2)}^N$,
the nuclear SSF $g_1^A$ can be evaluated in the quasi
elastic, inelastic and DIS regions.
Two different
prescriptions were used up to now to obtain the convolution formula:
the one of Ref. \cite{Bla} (to be called prescription 1),
corresponding to $C=0$
in Eq.\ (\ref{an})
(such a convolution formula has also been used in Ref.\ \cite{cps1}
where binding effects in
quasi--elastic scattering have been investigated), and the one
of
Ref.\ \cite{sa} (to be called
prescription 2)
corresponding to
$C=1$. The theoretical soundness of
both prescriptions, in particular some drawbacks
of prescription 1, as well as
their impact
on the quasi--elastic asymmetry, have been discussed in
Ref.\ \cite{sa} and in Ref.\ \cite{cps2}, and shall not be repeated here;
the important
point to be stressed, in the context of the present investigation,
is that in the Bjorken
limit  $(\nu /
| {\bf q} | \rightarrow 1$, $Q^2/| {\bf q} |^2 \rightarrow 0)$
both of them lead to the same result,
namely:
\be
g_1^A(x)  =  \sum_N \int _x ^A dz
{1 \over z} g_1^N \left( {x \over z} \right)
G^N(z)~~, \label{fin}
\ee
with the spin dependent light cone momentum distribution given by:
\widetext
\be
G^N(z)  =  \int dE\, \int d {\bf p}
\bigg \{  P_{||}^N( {\bf p},E )- \left[ 1 -
{p_{||} \over E_p + M} \right] {|{\bf p}| \over M}
{\cal P}^N({\bf p},E) \bigg\}
\delta \left(z - {p^+ \over M} \right)~~, \label{lux}
\ee
\narrowtext
where $p^+=p^0-p_{||}$ is the light cone momentum component.
We see that $g_1^A$ depends only upon $g_1^N$, whereas it
turns out \cite{cpss} that $g_2^A$ depends
both on $g_1^N$ and $g_2^N$.

In our calculations, the
nucleon SSF $g_1^N$
is the one proposed in Ref.
\cite{scha}, representing an extension of the Carlitz--Kaur model
\cite{car} by allowing
spin dilution of the valence quark due to
gluon polarization;the
effective nucleon polarization $p_{p(n)}$ are given by Eqs.\ (\ref{polp})
and (\ref{pol}); the
spin dependent spectral functions are the ones obtained in Ref.\cite{cps1},
yielding values of $p_{p(n)}$
(cf. Eqs.\ (\ref{up}) and (\ref{down})) in agreement with (\ref{polp})
and (\ref{pol}).

The $\vec{^3{\rm He}}$ asymmetry (Eq.\ (\ref{as}))
calculated using the convolution formula for $g_1^3$
(Eq.\ (\ref{fin})) and the
corresponding formula for
the unpolarized structure function
$F_2^3$ (see Ref.\cite{cps3}) is presented in Fig.\ref{fig1}(a),
and the
nuclear structure function $g_1^3$
is shown in Fig.\ref{fig1}(b).
The general trend of our results resembles the one found in
Ref.\cite{wol},
except for the asymmetry
at $x>0.9$ and $g_1^3$ at $x \simeq 0$.
We will discuss
the origin of these differences later on;
now we would like to stress the following point:
the non vanishing
proton contribution to the asymmetry shown in Fig.\ref{fig1}(a)
hinders in principle
the extraction of the neutron structure
function from the $\vec {^3{\rm He}}$ asymmetry.
As a matter of fact, once $g_1^3$ is obtained from the
experimental asymmetry, the
theoretically estimated proton contribution $g_1^{3,p}$ has to be
subtracted from it in order to obtain the neutron contribution
$g_1^{3,n}$. It can be seen from Fig.\ref{fig1}(b) that for
$0.01 \leq x \leq 0.3$ this quantity
differs from the neutron structure function $g_1^n$
by a factor of about $10\%$; since this factor is generated by
nuclear effects, one might be tempted to consider it
as the theoretical error on the determination of $g_1^n$;
however, it should be remembered that
the difference
between $g_1^n$ and $g_1^{3,n}$
is in principle model dependent
through
the way nuclear effects are introduced
and the specific form of $g_1^n$ used in the convolution
formula. Thus
it is necessary
to understand the origin of the nuclear effects
and how much they depend upon the form of $g_1^N$. To this end,
the
asymmetry and the structure function
predicted by the convolution approach are compared in Fig.\ref{fig2}
with the predictions of the simpler
models represented by Eqs.\ (\ref{gmod1})--(\ref{mod2}).
It can be
seen that
the model
which completely disregards nuclear effects
(binding and Fermi motion as well as $S'$ and $D$ waves),
predicts an asymmetry which strongly differs from the ones which
include nuclear effects; however it can also be seen that
at least for $x \leq 0.9$
nuclear effects can reliably be taken care of
by Eq.\ (\ref{mod2}), i.e. by considering
that
the only relevant nuclear effects are due to the effective nucleon
polarization induced by $S'$ and $D$ waves.
Such
a conclusion
is very clearly demonstrated in Fig.\ref{fig3}, where
the free neutron structure function is
compared with the quantity (cf. Eq.\ (\ref{gmod2})):
\be
\tilde g_1^n(x)={1 \over p_n} \left[ g_1^3(x)-2p_p g_1^p(x) \right]
\label{g1art}
\ee
calculated using the convolution formula for $g_1^3(x)$; it can be seen
that the two quantities are very close to each other, differing,
because of
binding and Fermi motion effects, by at most $4\%$.
Such a small difference
can be understood
by expanding ${1 \over z} g_1^N \left( {x \over z} \right)$
in Eq.\ (\ref{fin}) around $z=1$ and
by disregarding the term proportional to ${\cal P}^N$
in Eq. (16),
which gives anyway a very small contribution being of the order
$|{\bf p}| / M$; one obtains \cite{cpss}:
\be
g_1^{3,N}(x) \sim g_1^N(x) \left( p_N + {\Lambda_N \over M} \right)
+ x{d g_1^N(x) \over d x}
{\Lambda_N  \over M} + . . .
\label{tay}
\ee
where $\Lambda_N=\left[ \langle E_N \rangle ^{(+)} +
\langle T_{A-1}^N \rangle^{(+)} -
(\langle E_N \rangle^{(-)} +
\langle T_{A-1}^N \rangle^{(-)}) \right]$,
$\langle E_N \rangle^{(\pm)}$ and
$\langle T_{A-1}^N \rangle^{(\pm)}$ being the average removal and
recoil energies in the ``up" and ``down" states. Note that the
difference between these quantities appearing in $\Lambda_N$
results from the very definition of the polarized spectral function
$P_{||}^N$ (cf. Eq.\ (\ref{par}))(in unpolarized DIS, which is
governed by the unpolarized spectral function defined as the sum
of the ``up" and ``down" spectral functions,
the difference in $\Lambda_N$ is replaced by a sum \cite{cps3}).
Using the values of $\langle E_N \rangle^{(\pm)}$ and
$\langle T_{A-1}^N \rangle^{(\pm)}$ resulting from
three body realistic calculations \cite{cps1},
one gets $\Lambda_n/M  \sim 0.72 \cdot 10^{-3}$
and $\Lambda_p/M  \sim 0.25 \cdot 10^{-3}$, so
that the first term of Eq.\ (\ref{tay})
yields Eq.\ (\ref{gmod2}) and the second term,
representing Fermi motion and  binding corrections,
yields only a few percent contribution up to $x \sim 0.7$.
Thus we have theoretically
justified the correctness of Eq.\ (\ref{gmod2}) and
demonstrated that the
smallness of Fermi motion and binding is rather
independent of
the form of any well behaved $g_1^N$,
for large variations of $dg_1^N(x)/dx$ are killed
anyway by the smallness of $\Lambda_N$.
To sum up,
we have shown that the only relevant nuclear effects
in inclusive DIS of polarized
electrons off polarized $^3$He, are those related
to the proton and neutron effective polarizations arising
from $S'$ and $D$ waves, and that such a result does not crucially depend
upon the form of $g_1^N$.
Therefore, the
neutron structure function can be obtained from the $\vec{^3{\rm He}}$
asymmetry using in Eq.\ (\ref{g1art})
the
experimental values for $g_1^3$ and $g_1^p$ and the theoretical
quantities $p_p$ and $p_n$;
the resulting theoretical errors
due to Fermi motion and binding
(about 5$\%$) and to
the uncertainties
on $p_p$ and $p_n$ (cf. Eqs.\ (\ref{polp}) and (\ref{pol})),
lead to a total error
well below than hitherto assumed \cite{slac2,herm}.
The differences between our results and the ones of Ref.
\cite{wol} previously mentioned, are also clear:
the value of the
proton polarization generated by the wave function used in Ref. \cite{wol}
is $p_p=-0.023$, whereas our value is
$p_p=-0.030$, in full agreement
with Eq.\ (\ref{polp}).
It is therefore the combined effects of the underestimation
of the proton contribution and of the absence of binding effects,
which originate the shift upward of $g_1^3$ at $x \simeq 0$
and the flattening of $A_{\vec {^3{\rm He}}}$ at $x \sim 1$
exhibited by the results of Ref. \cite{wol} with respect to our ones.
In closing, we shall consider the first moment of the $^3$He
spin structure function, $viz$ $\Gamma_3=\int_0^1 g_1^3(x)dx$.
It can readily be shown that, provided the Bjorken sum rule \cite{bjo}
holds and the assumption (\ref{gmod2}) is valid, one has,
independently of the form of $g_1^{p(n)}$:
\be
\Gamma_3 & = & \int_0^1 g_1^3(x) dx= \nonumber \\
& = &  [p_n+2p_p]\Gamma_p-{ 1 \over 6 } { g_A \over g_V }
\left[ 1 - {\alpha_s \over \pi} \right]p_n
\label{bjo}
\ee
where $\Gamma_p=\int_0^1 g_1^p(x)dx$.
Using the values (\ref{polp}) and (\ref{pol}),
$g_A / g_V=1.259$ \cite{gav} and $\alpha_s=0$,
one gets $\Gamma_3=-0.180+0.804\Gamma_p$
(if the EMC result \cite{emc} is used for $\Gamma_p$
($\Gamma_p$=0.126), then $\Gamma_3=-0.079 \pm 0.003$,
the error being due to
the uncertainties on the values $p_p$ and $p_n$
(cf. Eqs. (\ref{polp}) and (\ref{pol})
(the error generated by
Fermi motion and binding is very small:
using the series expansion for $g_1^3$ and changing $\Lambda_N/M$
by a factor of 15, changes $\Gamma_3$ by less than 5$\%$).
The new experiments will provide both
$\Gamma_p$ and $\Gamma_3$, and the validity of
Eq.\ (\ref{bjo}) could be checked:
strong deviations of it
from the value $-0.180+0.804\Gamma_p$ can be
interpreted as evidence
of the violation of the Bjorken sum rule.
We have checked that various relativistic normalizations
of the spectral function affect neither $g_1^3$ nor $\Gamma_3$.

\eject
\begin{figure}
\caption{ (a) The $^3$He asymmetry (Eq.\ (\protect \ref{as}))
calculated within the convolution
approach (Eq.\ (\protect \ref{fin})(full)). Also shown are the
neutron (short dash) and proton (long dash) contributions.
(b) The SSF $g_1^3$ of $^3$He
(full); also shown are the neutron
(short dash) and proton (long dash) contributions. The dotted curve
represents the free neutron structure function
$g_1^n$. The difference between the dotted and short--dashed
lines is due to nuclear structure effects.}
\label{fig1}
\end{figure}

\begin{figure}
\caption{ (a) The $^3$He asymmetry
calculated with different nuclear
models. Dotted line: no nuclear effects (Eq.\ (\protect \ref{mod1}));
short-dashed line: $S'$ and $D$ waves
of $^3$He taken into account (Eq.\ (\protect \ref{mod2}));
long-dashed line: $S'$ and $D$ waves
of $^3$He taken into account plus Fermi motion effects;
full line: $S'$ and $D$ waves
of $^3$He taken into account plus
Fermi motion and
binding effects.
(b) The same as in (a) but for the SSF $g_1^3$ of
$^3$He. The long--dashed curve is hardly distinguishable
from the full one and it is not reported.}
\label{fig2}
\end{figure}

\begin{figure}
\caption{ The free neutron structure function $g_1^n$ (dots)
compared with the neutron structure function given by
Eq.\ (\protect \ref{g1art})(dash). The difference between
the two curves
is due to Fermi motion and binding effects.}
\label{fig3}
\end{figure}


\begin{references}
\bibitem[*]{scop}Part of this work was presented
at the University
of Perugia for partial
fullfillment of the degree of Dottore in Fisica .
\end{references}
\end{document}